\documentclass[aps,prl,reprint,twocolumn,superscriptaddress,showpacs,amssymb]{revtex4-1} 
\usepackage[english]{babel}
\usepackage{graphics}
\usepackage{graphicx}
\usepackage{epsfig}
\usepackage{amssymb}
\usepackage{dcolumn}
\usepackage{bm}
\usepackage{color}
\usepackage{lineno}
\usepackage{sidecap}
\usepackage{verbatim}  
\usepackage{vector}  
\usepackage{wrapfig}
\usepackage{enumerate}
\usepackage{sidecap}
\usepackage{amsmath}
\usepackage{hyperref}

\begin{document}

\title{Observation of Dirac surface states in the hexagonal PtBi$_2$, a possible origin of the linear magnetoresistance}

\author{S.\ Thirupathaiah}
\affiliation{Institute for Solid State Research, IFW Dresden, D-01171 Dresden, Germany.}
\affiliation{Solid State and Structural Chemistry Unit, Indian Institute of Science, Bangalore, Karnataka, 560012, India.}
\author{Y.\ Kushnirenko}
\affiliation{Institute for Solid State Research, IFW Dresden, D-01171 Dresden, Germany.}
\author{E.\ Haubold}
\affiliation{Institute for Solid State Research, IFW Dresden, D-01171 Dresden, Germany.}
\author{A. V.\ Fedorov}
\affiliation{Institute for Solid State Research, IFW Dresden, D-01171 Dresden, Germany.}
\author{E. D. L.\ Rienks}
\affiliation{Institute for Solid State Research, IFW Dresden, D-01171 Dresden, Germany.}
\author{T. K.\ Kim}
\affiliation{Diamond Light Source, Harwell Campus, Didcot, OX11 0DE, UK.}
\author{A. N.\ Yaresko}
\affiliation{Max-Planck-institute for Solid State Research, Heisenbergstrasse 1, 70569 Stuttgart, Germany.}
\author{C. G. F.\ Blum}
\affiliation{Institute for Solid State Research, IFW Dresden, D-01171 Dresden, Germany.}
\author{S.\ Aswartham}
\affiliation{Institute for Solid State Research, IFW Dresden, D-01171 Dresden, Germany.}
\author{B. B\"uchner}
\affiliation{Institute for Solid State Research, IFW Dresden, D-01171 Dresden, Germany.}
\author{S. V.\ Borisenko}
\affiliation{Institute for Solid State Research, IFW Dresden, D-01171 Dresden, Germany.}

\date{\today}

\begin{abstract}
The nonmagnetic compounds showing extremely large magnetoresistance are attracting a great deal of research interests due to their potential applications in the field of spintronics. PtBi$_2$ is one of such interesting compounds showing large linear magnetoresistance (MR) in its both the hexagonal and pyrite crystal structure. We use angle-resolved photoelectron spectroscopy (ARPES) and density functional theory (DFT) calculations  to understand the  mechanism of liner MR observed in the hexagonal PtBi$_2$. Our results uncover for the first time linear dispersive surface Dirac states at the $\bar{\Gamma}$-point, crossing Fermi level with node at a binding energy of $\approx$ 900 meV, in addition to the previously reported Dirac states at the $\bar{M}$-point in the same compound.  We further notice from our dichroic measurements that these surface states show an asymmetric spectral intensity when measured with left and right circularly polarized light, hinting at a substantial spin polarization of the bands.   Following these observations,  we suggest that  the linear dispersive Dirac states at the $\bar{\Gamma}$ and $\bar{M}$-points are likely to play a crucial role for the linear field dependent magnetoresistance recorded in this compound.
\end{abstract}

\maketitle
The property of magnetoresistance in the nonmagnetic solids has majorly two types of behaviour under the external magnetic fields. While the linear field dependent magnetoresistance as observed in graphene~\cite{Singh2012},  topological insulators Bi$_2$Te$_3$ and Bi$_2$Se$_3$ ~\cite{Qu2010, Wang2012b}, Ag$_{2+\delta}$Te/Se~\cite{Xu1997, Lee2002}, Dirac semimetals Cd$_3$As$_2$~\cite{Liang2014, Feng2015} and Na$_3$Bi~\cite{Kushwaha2015}, and type-I Weyl semimetals NbAs(P)~\cite{Ghimire2015, Shekhar2015, Wang2016a} and TaAs~\cite{Huang2015,Zhang2015a} is explained based on the linear dispersive Dirac states present near the Fermi level~\cite{Abrikosov1998, Abrikosov2003}, the quadratic field dependent MR as observed in the type-II Weyl semimetals WTe$_2$ and MoTe$_2$~\cite{Ali2014, Zhou2016}, LaSb~\cite{Zeng2016},  ZrSiS~\cite{Lv2016}, and Bi~\cite{Yang1999} is explained based on the charge compensation~\cite{Ali2014}. However, the role of charge compensation for the XMR of these compounds is still not clear as some reports recently demonstrated quadratic field dependent MR without charge balance~\cite{Zandt2007, Thirupathaiah2017}.   Moreover, the theory based on the impurity scattering or crystal disorder is inevitably brought into the above two cases and as well to the subquadratic field dependent MR~\cite{Fatemi2017, Parish2003, Ping2014}. The other existing mechanisms include metal-insulator transition~\cite{Khveshchenko2001, Kopelevich2006, Wang2014, Zhao2015, Xiang2015} and strong spin-orbit interactions~\cite{Jiang2015, Das2016}.  Thus, there is no clear consensus yet on the mechanism of magnetoresistance in the nonmagnetic solids.

PtBi$_2$ is a very interesting compound as it shows MR in both the hexagonal~\cite{Yang2016} and pyrite structures~\cite{Gao2016}. Astonishingly, the pyrite PtBi$_2$ has been found to show the highest MR observed among the known nonmagnetic metals so far~\cite{Xu1997, Lee2002, Liang2014, Ali2014, Feng2015, Shekhar2015, Zhou2016}, which also has been predicted as a 3-dimensional Dirac semimetal~\cite{Gibson2015}. Therefore, understating the mechanism of extremely large MR in this compound is essential due to its potential applications.  While in the pyrite structure the extremely large MR is attributed to its semimetalic nature~\cite{Gao2016}, the large linear MR observed in the hexagonal phase is explained based on the crystal disorder theory~\cite{Yang2016, Yao2016}.   Here, we report on the electronic structure of hexagonal PtBi$_2$ using the ARPES technique and first-principles calculations. Our studies suggest that PtBi$_2$ has two different surface terminations as the experimental band structure looks different from different cleavages. Our ARPES measurements demonstrate linear dispersive surface Dirac states near the Fermi level with a node at $\approx$ 150 meV below the Fermi level ($E_F$) at the $\bar{M}$-point with a 6-fold rotational symmetry in the $\overline{\Gamma M K}$ plane. Thus, there exists 6 Dirac cones of this type in a Brillouin zone. This observation is consistent with a previous ARPES report~\cite{Yao2016} on this compound.  Interestingly, in addition to this, we have uncovered for the first time an another surface Dirac cone at the zone center with a node at $\approx$ 900 meV below $E_F$. Much like in BiPd~\cite{Thirupathaiah2016}, the Dirac states at the zone center in PtBi$_2$ show a gap opening as a function of photon energy, possibly due to a surface-bulk hybridization.   Based on these studies,  we suggest that the linear MR observed in the hexagonal PtBi$_2$ could have originated from the linear dispersive surface states in the quantum limit, like in the case of topological insulators~\cite{Qu2010, Wang2012b} where the large linear MR is observed only from the surface band structure.

\begin{figure*}[t]
	\centering
		\includegraphics[width=0.98\textwidth]{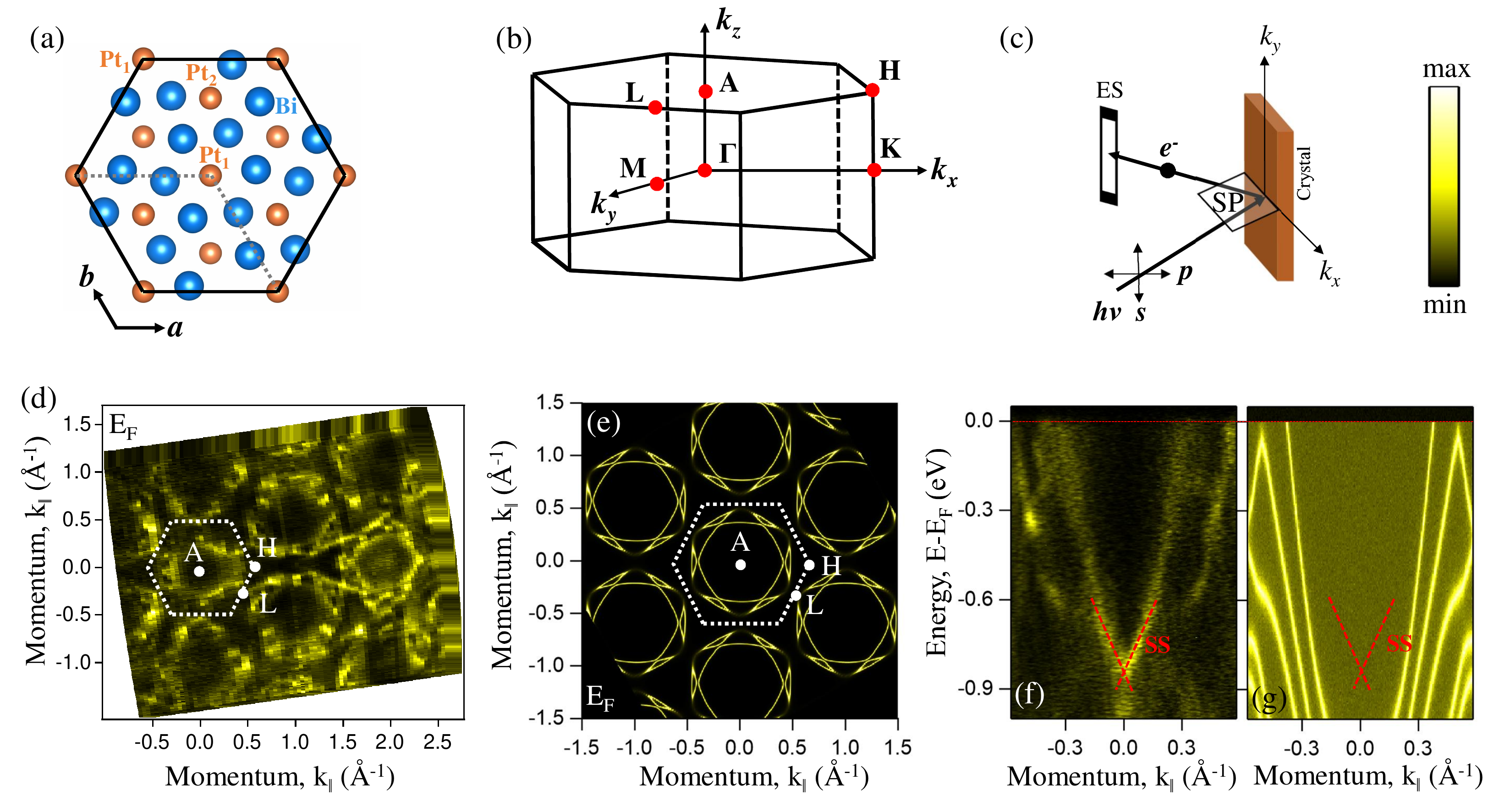}
	\caption{Electronic structure of PtBi$_2$. (a) Hexagonal crystal structure projected on to the $ab$-plane. (b) 3D view of the Brillouin zone with the high symmetry points. (c) Schematic representation of measuring geometry in which the $s$- and $p$-plane polarized lights are defined with respect to the analyzer entrance slit (ES) and the scattering plane (SP). (d) Fermi surface map extracted from the experiment using $p$-polarized light with a photon energy of 100 eV. (e) Fermi surface map derived from the DFT calculations. (f) Energy distribution map taken along the $A-L$ high symmetry line. (g) Energy-momentum ($E-k$) plot from the DFT calculations along the  $A-L$ high symmetry line. In the figure, the red dashed-lines represent the Dirac surface states.}
	\label{1}
\end{figure*}

 High quality single crystals of hexagonal PtBi$_2$ were grown using floating zone technique~\cite{Crystal}. The crystals have a platelet-like shape with shiny surface. The crystals were structurally characterized using powder X-ray diffraction to confirm the bulk purity and the hexagonal crystal structure with $P3$ space group.  ARPES measurements were carried out in BESSY II (Helmholtz Zentrum Berlin) synchrotron radiation facility at the UE112-PGM2b beam-line using the "1$^3$-ARPES"~\cite{Borisenko2012a,Borisenko2012b} end station equipped with SIENTA R4000 spectrometer.  The total energy resolution was set between  5 and 15 meV, depending on the incident photon energy. The measurements were performed at T = 1 K. Another set of ARPES measurements were performed in the Diamond light source at I05 beamline equipped with SIENTA R4000 spectrometer. The sample temperature at Diamond was always below 10 K and the energy resolution was set between 10 and 20 meV depending on the incident photon energy.

Self-consistent band structure calculations were performed on the hexagonal crystal structure of PtBi$_2$ having the lattice constants $a$ = $b$ = 6.572 \AA ~ and $c$ = 6.166 \AA~  using the linear muffin-tin orbital (LMTO) method~\cite{Andersen1975} in the atomic sphere approximation (ASA) as implemented in PY LMTO computer code~\cite{Antonov2004}. The Perdew-Wang parameterization~\cite{Perdew1992} was used to construct the exchange correlation potential in the local density approximation (LDA). Spin-orbit coupling was taken into account at the variational step.

\begin{figure*}
	\centering
		\includegraphics[width=0.98\textwidth]{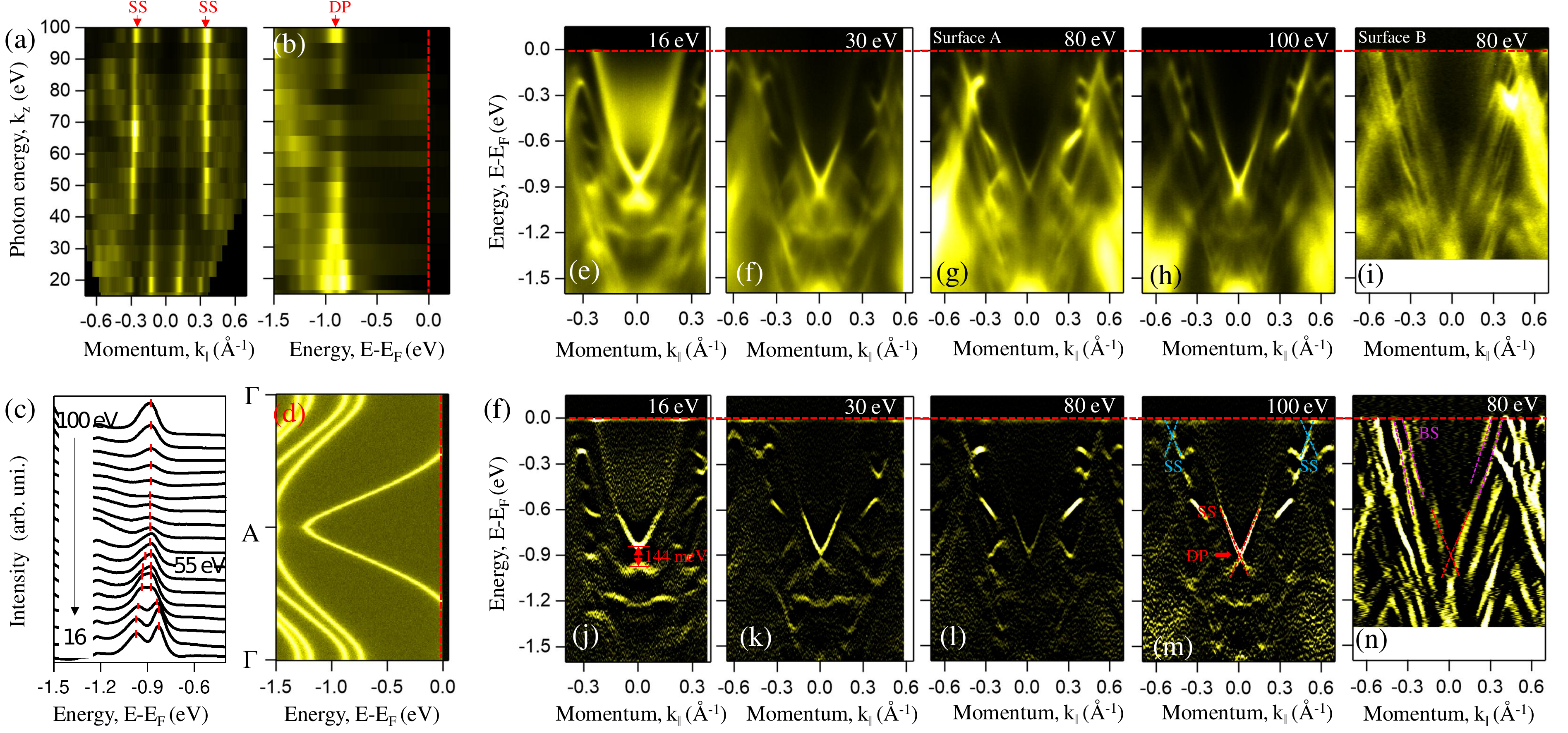}
	\caption{Photon energy dependent ARPES data. The data are measured using $p$-polarized light. (a) $k_z$ Fermi surface map taken in the $\Gamma M L A$ plane. (b) Energy distribution map taken along the $\Gamma-A$ symmetry line at $k_y$, $k_x$=0. (c) Energy distribution curves as a function of photon energy. (d) $E-k$ plot derived from the DFT calculations in the $\Gamma$-$A$ high symmetry line. (e)-(h) Photon energy dependent EDMs taken along the $\Gamma$-$M$ direction. (j)-(m) second derivatives of (e)-(h). (i) EDM measured using $p$-polarized light along the $\Gamma$-$M$ symmetry line with a photon energy of 80 eV from a different cleavage plane than the rest of the data. (n) second derivative of (i). In the figure, the red and blue dashed lines represent the surface states and the pink dashed lines represent the bulk states. Dirac point (DP) of the surface states at the $\bar{\Gamma}$-point is at around 900 meV and at the $\bar{M}$-point is at around 150 meV below $E_F$.}
	\label{2}
\end{figure*}

The electronic structure of PtBi$_2$ is shown in Fig.~\ref{1} obtained by means of ARPES and DFT calculations.  The ARPES data are measured using $p$-polarized light with a photon energy of 100 eV. The data are taken at a sample temperature of 1 K.  We estimate that 100 eV photons detect the bands nearly from the $k_z$=5.5 $\pi/c$ plane using an inner potential of Pt of 25 eV~\cite{Ichikawa2003}, i.e., from the $A L H$ plane. Fig.~\ref{1}(a) shows the hexagonal crystal structure of PtBi$_2$ projected onto the $ab$-plane. Fig.~\ref{1}(b) shows 3D view of the Brillouin zone in which various high symmetry points are located.  In Fig.~\ref{1}(c) we  schematically show the measuring geometry where the $s$- and $p$-plane polarized lights are defined with respect to the analyzer entrance slit (ES) and the scattering plane (SP). Fig.~\ref{1}(d) shows the experimental Fermi surface (FS) map in the $k_x-k_y$ plane extracted by integrating over an energy window of 20 meV centered at $E_F$. Fig.~\ref{1}(f) shows  the FS map derived from DFT calculations.  Fermi surface obtained in our studies [see Fig.~\ref{1} (d) and (e)] is inline with the hexagonal crystal structure of PtBi$_2$[see Fig.~\ref{1}(a)]. Fermi surface obtained in our studies is inline with the hexagonal crystal structure of PtBi$_2$ [see Fig.~\ref{1}(a)]. Energy distribution map (EDMs) along the $A$-$L$ high symmetry
lines is shown in Fig.~\ref{1} (h). In Fig.~\ref{1} (h), the red-dashed lines represent linearly dispersive Dirac surface states with a node at $\approx$ 900 meV. Calculated band dispersions along the $A$-$L$ high symmetry lines are shown in Fig.~\ref{1} (i). Note here that the Fermi level of the calculated band structure is shifted about 200 meV towards the lower-binding energy to match with the experiment. A similar observation was also made in the previous ARPES report of this compound~\cite{Yao2016}.




\begin{figure} [ht]
	\centering
		\includegraphics[width=0.48\textwidth]{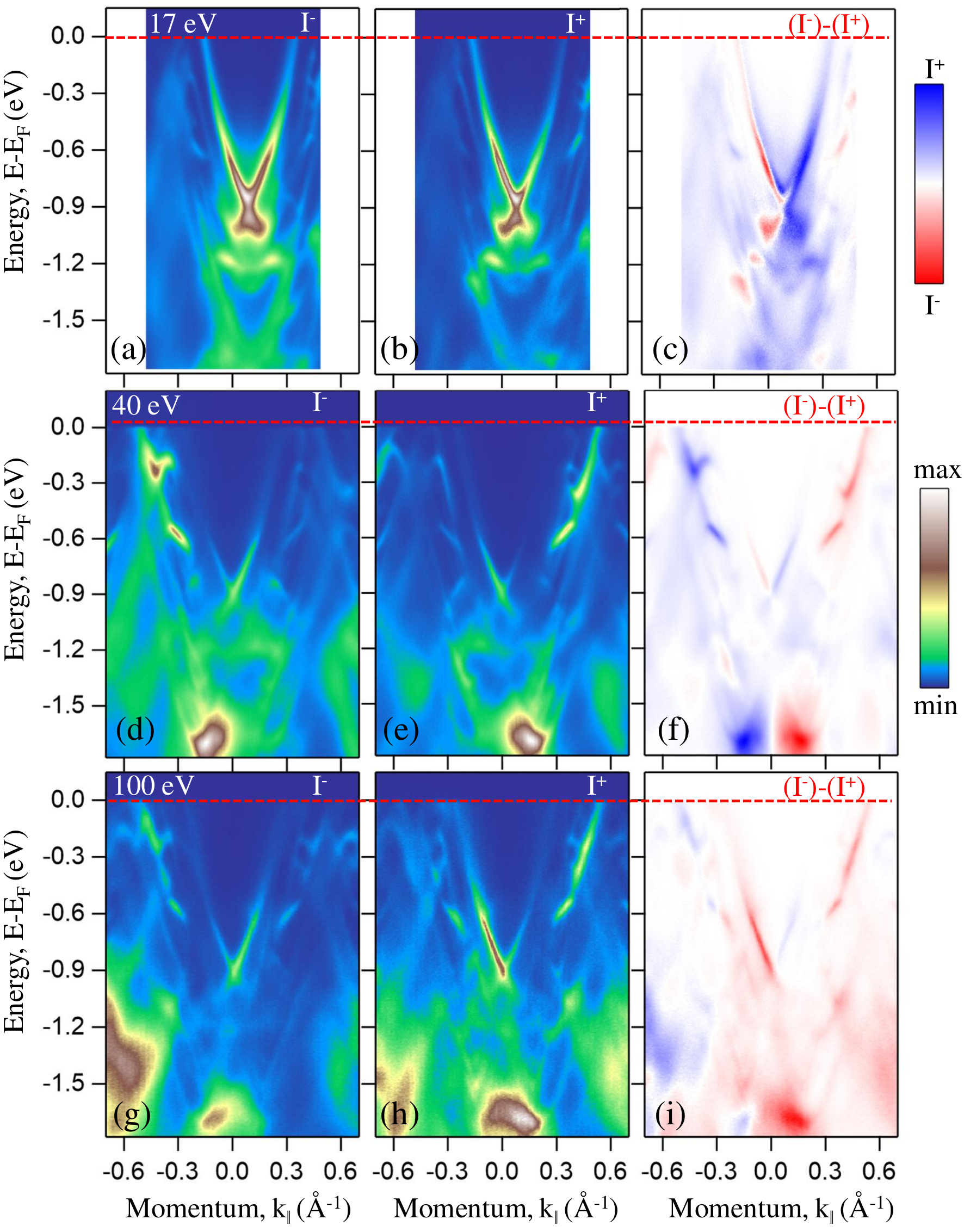}
	\caption{Circular dichroism.  (a) EDM measured using left circular ($I^-$),  (b) right circular ($I^+$) and (c) the resultant EDM after the spectral intensity subtraction ($I^-$-($I^+$) measured with a photon energy of 17 eV. (d)-(i) Similar data of (a)-(c) but measured with 40 and 100 eV photon energies.}
	\label{3}
\end{figure}

Next, we show photon energy dependent ARPES measurements performed to understand the $k_z$ dependent band structure of this compound. Fig.~\ref{2}(a) shows the $k_z$ Fermi surface map measured using $p$-polarized light in the $\Gamma M L A$ plane. Fig.~\ref{2}(b) shows the EDM taken along the $\Gamma$-$A$ high symmetry line at $k_x, ~k_y$=0. Energy distribution curves (EDCs) as a function of incident photon energy are shown in Fig.~\ref{2}(c). Band dispersions in the $\Gamma$-$A$ symmetry line derived from the DFT calculations are shown in Fig.~\ref{2}(d). Figs.~\ref{2}(e)-(h) depict EDMs measured using various photon energies. Figs.~\ref{2}(j)-(m) depict second derivatives of Figs.~\ref{2}(e)-(h). In Fig.~\ref{2}(i), we show EDM taken from a different cleavage (say Surface B) surface using $p$-polarized light with a photon energy of 80 eV, while the rest of the data are measured from another cleavage (say Surface A).  Fig.~\ref{2}(n) depicts the second derivative of Fig.~\ref{2}(i). In the figures, the red- and blue-dashed lines represent the surface states (SS), while the pink-dashed lines represent the bulk states (BS).

In the $k_z$ map shown in Fig.~\ref{2}(a), we can clearly notice that the two Fermi sheets shown by the red down-arrows have negligible $k_z$ warping. In principle, surface states show $k_z$ independent band structure. On the other hand, our bulk band structure calculations suggest a strong $k_z$ warping of the Fermi sheets in the $\Gamma M L A$ plane (not shown) . Therefore, from these observations we suggest that these Fermi sheets are from the surface contribution. This is further confirmed from the EDM taken along the $\Gamma$-$A$ direction [see Fig.~\ref{2}(b)] where one can notice a flat dispersion for the Dirac point (DP) at $\approx$ 900 meV of the binding energy and no such band is found from a similar EDM  derived from the calculations [see Fig.~\ref{2}(d)]. Interestingly, from the energy distribution curves (EDCs) extracted as a function of photon energy we noticed that the Dirac states at the zone center are gapped when measured with the photon energy $h\nu$ $<$ 55 eV. This is further demonstrated from the photon energy dependent EDMs shown in Figs.~\ref{2}(e)-(h) and from the corresponding second derivatives shown in Figs.~\ref{2}(j)-(m).   The estimated maximum energy gap is of 145 meV when measured with 16 eV of photon energy [see Fig.~\ref{2}(j)]. A similar observation was made earlier in the case of noncentrosymmetric BiPd~\cite{Thirupathaiah2016}. There, it was suggested that the gap opening is due to the hybridization between the surface and the bulk states located in the same binding energy range. In the present case as well it could be the similar scenario as the bulk bands disperse in the $k_z$ direction over a window of 1 eV of binding energy. As can be seen in Fig.~\ref{2}(d), the binding energy of the band at $\Gamma$ is of 0.65 eV whereas at $A$ the binding energy of the same band is of 1.5 eV. Therefore, as the surface states show a finite penetration depth into the solid~\cite{Autes2015, Li2009j}, it is highly likely to have a surface-bulk hybridization which may lead to the gap opening at the Dirac point. On Fig.~\ref{2}(m), we have shown schematically another set of Dirac surface states  by the  blue-dashed lines located at the $\bar{M}$-point having a Dirac point at $\approx$ 150 meV with a 6-fold rotational symmetry. Thus, there are in total 6 Dirac cones of this type in a BZ. This observation is in good agreement with previous ARPES report on this compound~\cite{Yao2016}. The electronic structure of PtBi$_2$ seems to be cleavage dependent. As can be seen in Figs.~\ref{2}(i) and (n), the EDM taken from the cleavage surface B shows no signatures of the Dirac-like surface states when compared to the data taken on the cleavage surface A. Moreover, the band structure highlighted by pink-dashed lines shown in Figs.~\ref{2}(i) and (n) look much alike the bulk band structure derived from the DFT calculations [see Fig.~\ref{1}(i)].



Next we show the circular dichroic ARPES measurements.  Energy distribution map (I$^-$)  shown in Fig.~\ref{3}(a) is measured using a left-circular polarized light with a photon energy of 17 eV.  Similarly, EDM (I$^+$) shown in Fig.~\ref{3}(b) is measured using a right-circular polarized light. Fig.~\ref{3}(c) is the resultant spectra after the subtraction (I$^-$-I$^+$). Figs.~\ref{3}(d)-(i) are analogous data to Figs.~\ref{3}(a)-(c) but measured with photon energies of 40 and 100 eV. Generally, in spin-orbit coupled systems the helicity of light can couple to the electron spin degree of freedom through the total angular momentum. Therefore, the helicity dependent spectral intensity can provide indirectly information on the spin polarization of the detected sates~\cite{Wang2011,Park2011, Wang2013}. However, it is also argued that such measurements suffer greatly from the final state effects and thus, reversing the dichroic signal among the photon energies~\cite{Scholz2013}. Nevertheless, as can be seen from the intensity difference spectra [Figs.~\ref{3}(c), (f) and (i)] taken from a wide range of photon energies (17, 40 and 100 eV), the dichroic signal is consistent at the zone center switching the spectral intensity of the Dirac cone as $k\rightarrow-k$. Interestingly, dichroic signal is noticed also for the Dirac states at the $\bar{M}$-point,  but with opposite sign.

Previous transport~\cite{Yang2016} and ARPES studies~\cite{Yao2016} on this compound suggested crystal disorder for the observed linear magnetoresistance. On the other hand, our present studies clearly show two sets of linear dispersive Dirac states crossing the Fermi level with nodes located at around 150 meV and 900 meV below $E_F$. Therefore, based on these observations we suggest that the linear MR observed in this compound could be originated from the linear dispersive surface states.  This is argument is inline with the theory  based on the Landau levels of the linear dispersive states in the presence of external magnetic field~\cite{Abrikosov1998, Abrikosov2003}, where it was suggested that the linear dispersive states without a gap near the Fermi level play a crucial role. In addition, following our circular dichroic measurements we further suggest that the surface states are substantially spin polarized. Therefore, spin polarization of the bands also may play a vital role for the magnetoresistance~\cite{Jiang2015} of this compound.

In conclusion, we studied the low-energy band structure of the hexagonal PtBi$_2$ by means of ARPES technique and DFT calculations.  Our studies uncover for the first time linearly dispersive Dirac states near $\bar{\Gamma}$-point, crossing the Fermi level with a node at $\approx$ 900 meV below $E_F$. In addition, another surface Dirac cone is found at the $\bar{M}$-point-point with a node at $\approx$ 150 meV below $E_F$. We further notice from the circular dichroic measurements that these surface states show a significant dichroic signal switching as $k\rightarrow-k$. Following these studies, we suggest that the linear dispersive spin polarized Dirac states found at both the $\bar{\Gamma}$-point and $\bar{M}$-point points could be a possible origin for the large linear MR observed in this compound rather than the crystal disorder that was suggested earlier for the same.

This work was supported	under the DFG grant BO 1912/7-1. S.T.  acknowledges support by the Department of Science and Technology, India through the INSPIRE-Faculty program (Grant number: IFA14 PH-86). We acknowledge Diamond Light Source for time on Beamline/Lab I05 under Proposal SI15936-1.
\bibliography{PtBi2}

\begin{thebibliography}{50}
\bibitem [{Cry()}]{Crystal}%
  \BibitemOpen
  \href@noop {} {\enquote {\bibinfo {title} {{Stoichiometric amounts of Pt and
  Bi were alloyed using an arc-melter, then crushed and filled in a quartz
  crucible with 4mm inner diameter. The filled crucible was moved fast through
  the zone to create a dense filling. Afterwards the filled crucible was zone
  refined without sample rotation with a speed of 5mm/h. The resulting ingot
  was characterized with powder XRD of selected pieces. The Rietveld method was
  used to derive lattice parameters using the fullprof program. By EDX-analysis
  the estimated composition is PtBi$_2$ and is stoichiometric with in the error
  bars of EDX.}}}\ }\BibitemShut {NoStop}%
\end{thebibliography}

\end{document}